\documentclass[pra,10pt,twocolumn]{revtex4}
\usepackage[dvips]{epsfig}
\usepackage{float}
\usepackage{upgreek}
\usepackage{amsfonts}
\usepackage{amsmath}
\usepackage{amssymb}
\usepackage{enumerate}
\usepackage{pslatex}
\usepackage{hhline}
\usepackage{threeparttable}
\usepackage{supertabular}
\usepackage{multirow}
\usepackage{tabularx}

\newcommand{\e}{\varepsilon}

\newcommand{\vJ}{\mathbf{J}}

\newcommand{\ve}{\mathbf{e}}

\newcommand{\vn}{\boldsymbol{\nabla}}

\newcommand{\braket}[2]{\langle #1|#2\rangle}
\newcommand{\vE}{\mathbf{E}}
\newcommand{\vrr}{\mathbf{r}}
\newcommand{\vmu}{\bm{\mu}}
\newcommand{\vecb}[1]{\mathbf{#1}}

\newcolumntype{Y}{>{\centering\arraybackslash}X}

\begin{document}

\title{Coherent interference effects in a nano-assembled optical cavity-QED system}

\author{Paul E. Barclay}
\email{paul.barclay@hp.com} \altaffiliation{Presently with Hewlett-Packard Laboratories, 1501 Page Mill Rd, Palo Alto, CA 94304, USA}
\affiliation{Thomas J. Watson, Sr.,\ Laboratory of Applied Physics, California Institute of Technology, Pasadena, CA 91125, USA}
\author{Oskar Painter}
\email{opainter@caltech.edu}
\homepage{http://copilot.caltech.edu}
\affiliation{Thomas J. Watson, Sr.,\ Laboratory of Applied Physics, California Institute of Technology, Pasadena, CA 91125, USA}
\author{Charles Santori}
\author{Kai-Mei Fu}
\author{Raymond G. Beausoleil}
\affiliation{Hewlett-Packard Laboratories, 1501 Page Mill Rd, Palo Alto, CA 94304, USA}
\date{\today}

\maketitle

\setcounter{page}{1}

\noindent

\textbf{Fano resonances\cite{ref:Fano1}, the signature of multi-path interference between a continuum and discrete resonance, have been
exploited in optics\cite{ref:Wood1,ref:Fan5} for a variety of applications including biosensing\cite{ref:Chao1}, optical
switching\cite{ref:Cowan3} and wavelength conversion\cite{ref:Mondia1}.  In cavity-QED\cite{ref:Mabuchi}, photons stored in a cavity leak into a
continuum of radiation modes along with the atomic spontaneous emission. Measurements of this radiation can, in principle, yield information
regarding the strength of atom-cavity interactions\cite{ref:Carmichael4,ref:Harbers1}.  Utilizing nano-assembly techniques to integrate the
constituent components of a solid-state cavity-QED system, here we realize a platform for studying interference phenomena of an emitter coupled
to a microcavity and its radiation mode environment.  A quantum model of the system is presented, from which the coherent coupling rate between
cavity and emitter is estimated.  It is envisioned that this nano-assembly approach may be applied to the integration of more complex cavity-QED
geometries, where the optical coupling and entanglement of multiple quantum emitters can be realized.}

\begin{figure}[b]
\begin{center}
  \epsfig{figure=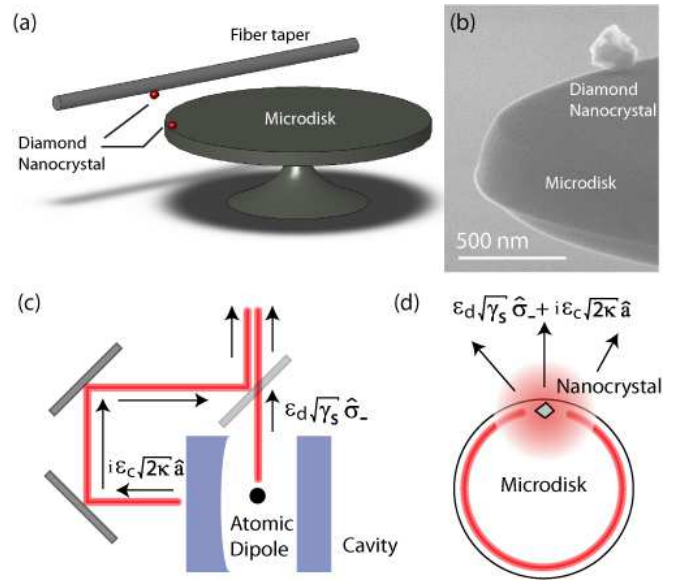, width=1\linewidth}
\caption{(a) Schematic of the nanocrystal fiber-to-microcavity positioning technique. (b) SEM image of a diamond
  nanocrystal positioned on the edge of a SiO$_2$ microdisk.  Illustration of the interference between direct dipole and indirect dipole-cavity radiation, (c) for a generic Fabry-Perot
  single-sided cavity and (d) for a microdisk cavity with nanocrystal-scattering radiation loss.}\label{fig:schematic_SEM}
\end{center}
\end{figure}

The cavity QED system studied in this Letter (Fig.\ \ref{fig:schematic_SEM}) consists of a diamond nanocrystal containing optically active
nitrogen-vacancy (NV) centers that is coupled to a dielectric microdisk cavity.  Optical access to the nanocrystal-microdisk system is provided
by both an optical fiber taper waveguide\cite{ref:Knight} and a high numerical aperture (NA=0.55) objective lens.  In bulk diamond, NV centers
exhibit atom-like optical properties, and are promising candidates for applications in quantum information processing, including single photon
generation\cite{ref:Gruber1}, coherent population trapping\cite{ref:Santori2}, and optical readout and manipulation of single nuclear
spins\cite{ref:Jelezko1,ref:Dutt1,ref:Childress1}. NV centers hosted in diamond nanocrystals\cite{ref:Beveratos2} are amenable to
nanomanipulation\cite{ref:Sandoghdar1} and integration with nanophotonic structures\cite{ref:Park1,ref:Wang2}. The fiber taper waveguide is a
versatile tool in this regard; in addition to serving its usual role as a probe for the optical fields of the microcavity, in this work it also
functions as a collection optic and a positioning tool for diamond nanocrystals.  This provides several benefits, notably pre-selection of
nanocrystals with optimal spectral properties from a nominally lower quality ensemble\cite{ref:Shen1}, and the controlled positioning of any
number of nanocrystals on photonic structures.

Light radiated from a coupled emitter (hereafter referred to as a ``dipole'') and cavity system is usually described by two, distinct, source
channels.  The cavity radiation channel consists of the localized quasi-mode of the cavity and the radiation modes which it leaks into, whereas
the dipole radiation channel consists of all other radiation modes directly coupled to the dipole emitter.  In the system studied here, the
radiation mode emission from the cavity mode and the dipole overlap, as illustrated for a generic Fabry-Perot cavity in Fig.\
\ref{fig:schematic_SEM}(c), resulting in interference between the radiated field from each channel. A quantitative understanding of the
interference effects can be obtained through a quantum mechanical model of the dipole-cavity system.  For simplicity we consider a single mode
of the microdisk which leaks into radiation modes with energy decay rate $2\kappa$.  The NV center optical transition is modeled by a single
dipole transition, with excited state energy decay rate due to spontaneous emission $\gamma_s$ and pure dephasing rate $\gamma_p$. The
coherent-coupling rate between the dipole transition and an excitation of the microcavity mode is $g$.  The electric field radiated into the
far-field by the dipole-cavity system can then be written as,

\begin{equation}\label{eq:field}
  \hat{\vE}(\vrr,t) = \mathbf{e}_d(\vrr)e^{-i\phi_d}\sqrt{\gamma_s}\hat{\sigma}_-(t) + \mathbf{e}_c(\vrr)e^{-i\phi_c}
  \sqrt{2\kappa}~\hat{a}(t)+\text{h.c.},
\end{equation}

\noindent where $\hat{a}$ is the microcavity field operator and $\hat\sigma_-$ is the polarization operator of the dipole transition.  In
general, the field profiles $\mathbf{e}_d(\vrr)$ and $\mathbf{e}_c(\vrr)$, of radiation from the dipole and the microcavity, respectively, need
not be orthogonal.  The phases $\phi_{d,c}$ associated with the field operators $\hat{a}$ and $\hat{\sigma}_-$ depend on the system and
measurement geometry. In the generic example drawn in Fig.\ \ref{fig:schematic_SEM}(c), the relative phase between the direct and indirect
emission is determined by the additional path length followed by the cavity emission, and by the phase imparted by the cavity output coupling.
For the nanocrystal-microdisk system (Fig.\ \ref{fig:schematic_SEM}(d)), the path length difference can be zero.

The time evolution of the operators $\hat{\sigma}_-$ and $\hat{a}$ can be calculated from the system density matrix equation of motion, which
depends on the dipole-cavity Hamiltonian and on Lindblad operators representing the cavity and dipole decoherence
processes\cite{ref:Carmichael4} (see Appendix).  In the ``room temperature'' limit, $\gamma_p \gg \kappa,\gamma_s,g,\Delta\omega$, and the
detected optical spectra into a given collection optic is:

\begin{equation}\label{eq:spectra}
S(\omega) = \frac{1}{\gamma_p}\left| \upepsilon_d e^{-i\phi_d} + \upepsilon_c
  e^{-i\phi_c}\sqrt{\frac{2g^2}{\kappa\gamma_s}}\frac{1}{1+i\Delta\omega/\kappa}\right|^2,
\end{equation}

\noindent where $\Delta\omega = \omega_c - \omega$ is the cavity-dipole emission  detuning, $\upepsilon_{d,c} =
\braket{\mathbf{e}_{o}(\vrr)}{\mathbf{e}_{d,c}(\vrr)}_\vrr$ describes the overlap between the mode $\mathbf{e}_o(\vrr)$ of the collection optic
and that of the dipole (cavity) radiation.  Note that the relative amplitude of the cavity emission scales with $F_o = 2g^2/\kappa \gamma_s$,
the bad-cavity Purcell factor.  Also note that in the room temperature limit, the effect of phonon-assisted emission on the dipole-cavity
dynamics can be included as a dominant contribution to $\gamma_p$.

\begin{figure}[tb]
\begin{center}
  \epsfig{figure=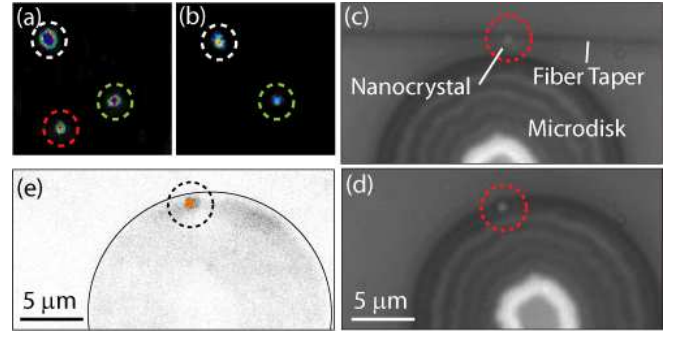, width=1\linewidth}
  \caption{Scanning confocal microscope (SCM) images (details in \cite{ref:Santori2}) of a mesa before (a) and after (b)
    a nanocrystal has been picked up with a fiber taper.  Optical images of (c) fiber taper and attached nanocrystal
    aligned with microdisk edge, and (d) after nanocrystal placement. Nanocrystal imaging is aided by a white light
    source coupled into the fiber taper and the microdisk.  (e) SCM image of microdisk after nanocrystal placement.} \label{fig:pick_place_images}
\end{center}
\end{figure}

\begin{figure}[tb]
\begin{center}
  \epsfig{figure=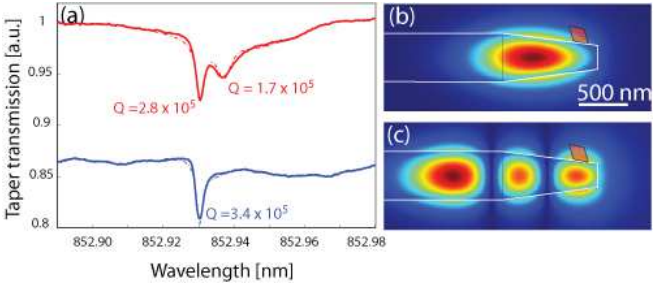, width=1\linewidth}
  \caption{(a) Microdisk mode lineshapes measured by monitoring the fiber taper transmission spectrum with the taper positioned in the microdisk near-field,
   before (blue) and after (red) nanocrystal placement.  A 850nm band tunable diode laser source (New Focus Velocity) was used to measure the taper transmission; a 630nm tunable source was not readily
    available at the time of the measurements. The dashed lines are fits \cite{ref:Borselli2} to the resonant modes.  Cross-sectional view of the $630$ nm wavelength band (b) TM$_{p=1}$ and (c) TM$_{p=3}$ whispering-gallery mode profiles near the disk edge as calculated using the finite-element-method (FEM).  The white outline indicates the periphery of the microdisk, with the shape of the disk profile estimated from SEM images.  Only the dominant, vertical component of the electric field is plotted for clarity.  The position of the nanocrystal, as placed on the
disk, is indicated by the red diamond.} \label{fig:WGM_properties}
\end{center}
\end{figure}

The microdisk cavities studied in this work are $20\mu$m in diameter, $400$ nm thick, and are formed by thermal oxidization of pre-patterned Si
microdisks\cite{ref:Borselli2}.  The resulting SiO$_2$ microdisks (refractive index $n_{\text{SiO}_{2}} \approx 1.45$) support high-$Q$
whispering-gallery-modes (WGMs) across the visible and near-infrared spectrum, with measured values as high as $Q = 8 \times 10^5$ in the
$830$-$855$ nm wavelength band.  Of particular interest are the transverse-magnetic-like (TM-like) disk modes, which are polarized primarily
normal to the planar microdisk resulting in significant electric field intensity at and above the disk surface (Fig.
\ref{fig:WGM_properties}(b,c)).  For a wavelength of $637$ nm, near the emission wavelength of the negatively charged NV transition (NV$^-$),
the effective mode volume of the lowest radial-order ($p=1$) TM-like WGM (TM$_{p=1}$) is calculated to be $V_{\text{eff}} =
82(\lambda/n_{\text{SiO}_{2}})^3$. The maximum ratio between the electric field energy density at the disk surface to that at the point of peak
electric field energy density (lying within the disk) is $1/\eta=0.23$.  For the NV$^-$ transition of a nanocrystal placed at this optimal
surface location, and with dipole orientation aligned normal to the disk surface, this translates into a coherent coupling rate between the
TM$_{p=1}$ mode and the NV$^-$ dipole of $g_{\text{o}}/2\pi \approx 0.64$ GHz.  This estimate is based upon a total excited state spontaneous
emission rate of $\gamma_{||}/2\pi \approx 12$ MHz\cite{ref:Tamarat1}.  The situation, however, is complicated for the NV$^-$ transition due to
electron-phonon interactions which result in signigicant coupling to several phonon sidebands.  Of particular interest is coupling to the NV$^-$
zero phonon line (ZPL), which for the $3$-$5\%$ fraction of spontaneous emission that is emitted \emph{into the
ZPL}\cite{ref:Manson1,ref:Davies2}, yields a reduced coherent coupling rate to the TM$_{p=1}$ of $g_{\text{zpl}}/2\pi \approx 0.13$ GHz.

The nanopositioning scheme used to assemble individual diamond nanocrystals on the surface of the microdisks is illustrated in Fig.\
\ref{fig:schematic_SEM}(a).  A nanoscale taper formed from single mode silica fiber (Nufern 630-HP) is used both as a collection optic and as a
means to pick-and-place diamond nanocrystals.  The diamond nanocrystals (Diamond Innovation-NAT, 200nm median diameter) are initially sparsely
dispersed on a clean silicon sample patterned with elevated mesa structures.  The diamond coated silicon sample, the microdisk sample, and the
optical fiber taper are mounted in an enclosed box with a dry nitrogen purge. After identifying a nanocrystal of
interest\cite{ref:Santori2,ref:Shen1}, the fiber taper is contacted on top of a nanocrystal using high resolution (50 nm) stages.  The taper is
then raised vertically away from the mesa surface. Due to surface interactions, which are not yet fully understood, the nanocrystal attaches to
the fiber taper and is lifted from the silicon mesa surface with high yield (see Figure \ref{fig:pick_place_images}(a-b)).  Transferring the
nanocrystal to a microcavity follows the ``pick-up'' process in reverse (Fig.\ \ref{fig:pick_place_images}(c-d)).  In order to detach the
nanocrystal from the fiber taper it was found necessary to slightly move the microdisk ($1-10$ $\mu$m in-plane steps) when in contact with the
nanocrystal so as to rub it off of the taper.  Images of the microdisk after placement of the nanocrystal ($\approx 200$ nm diameter) are shown
in Figs. \ref{fig:schematic_SEM}(b) and Fig.  \ref{fig:pick_place_images}(e). Using this technique nanocrystals can typically be placed within a
taper diameter ($ < 500$ nm) of the disk edge.  In this case, the nanocrystal has been placed approximately $150$ nm from the disk edge.  To
facilitate widefield optical imaging of the nanocrystal during positioning, a relatively large particle was chosen in this instance; smaller
nanocrystals can also be manipulated and positioned appropriately with the aid of confocal microscopy to identify the nanocrystal position.

\begin{figure}[tb]
\begin{center}
  \epsfig{figure=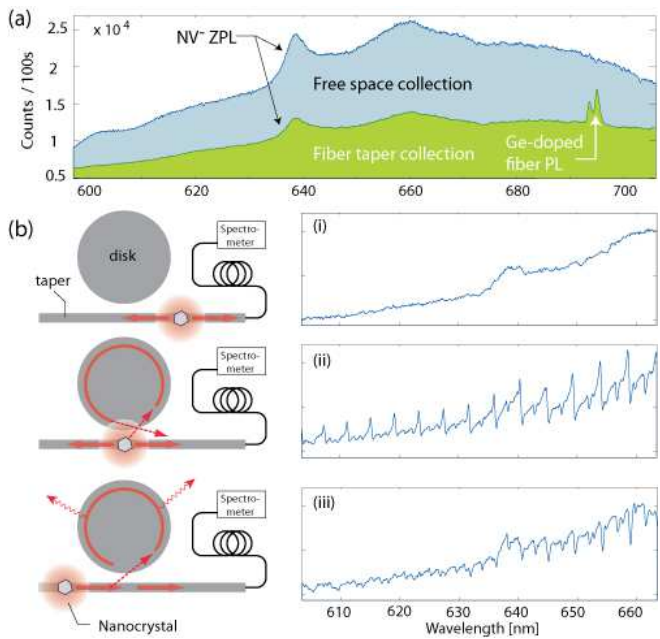, width=1\linewidth}
  \caption{(a) Emission from a diamond nanocrystal attached to a fiber taper, collected through the high-NA lens in the far-field and through the
    fiber taper in the near-field.  The fiber taper data was scaled by a factor of 1.6 to take measured fiber insertion loss into
    account. Peaks near $\lambda=690$ nm are due to fluorescence in the Ge doped fiber core.  (b) Measured emission when
    the taper interacts with a microdisk, for varying nanocrystal position relative to the microdisk and spectrometer,
    as indicated by the illustrations.}\label{fig:apparatus_taper_data}
\end{center}
\end{figure}

At each step in this pick-and-place process we use the fiber taper, in combination with widefield imaging optics, to study the nanocrystal
emission. Figure \ref{fig:apparatus_taper_data}(a) shows the photoluminescence (PL) spectra after the first step in which the diamond
nanocrystal is attached to the fiber taper.  Emission is collected both from a high-NA objective lens at normal incidence to the sample as well
as through the fiber taper.  Optical excitation is performed using a frequency-doubled YAG laser ($\lambda \approx 532$ nm).  The pump laser beam
is sent through the collection-lens and focused down to a 1.5 $\mu$m spot, with a resulting pump intensity of $\sim 10$ kW/cm$^2$.  Emission
from the NV$^-$ ZPL is visible at $\lambda \approx 637$nm, superimposed on the large phonon sideband characteristic of diamond NV centers.  The
measured efficiency of the fiber taper collection relative to the lens collection is $\sim 40\%$.  Theoretically, using smaller diameter fiber
tapers ($\sim 300$nm) and nanocrystals ($\sim 50$nm), it should be possible to reach an absolute taper collection efficiency of $ > 10\%$
\cite{ref:Klimov1,ref:LeKien1}.

Figure \ref{fig:apparatus_taper_data}(b) shows measured PL spectra during the next step in the process in which the nanocrystal is brought into
the near-field of the microdisk using the fiber taper.  In this step the fiber taper waveguide is aligned and evanescently coupled to the
microdisk with the nanocrystal positioned (i) far from the microdisk, on the nearside of the spectrometer input, (ii) nearly touching the
microdisk in the taper-microdisk coupling region, and (iii) far from the microdisk, on the far side of the spectrometer input.  In (i) the
microdisk has no significant effect on the measured NV emission spectrum as only light directly emitted into the fiber reaches the spectrometer.
In (iii) the microdisk behaves as a simple drop filter on the emission radiated into the fiber, resulting in Lorentzian dips at each of the
cavity resonances.  The spectrum in (ii) is more complex, with various Fano lineshapes appearing in the spectrum due to interference between the
cavity and taper spontaneous emission channels (more about this below).

In the final step of the nanocrystal-microdisk assembly process, the nanocrystal is transferred onto the surface of the microdisk.  The strong
interaction between the nanocrystal and microdisk WGMs is evidenced by the effect of the nanocrystal on the WGM spectral properties. Prior to
placement of the nanocrystal, the microdisk in Fig.\ \ref{fig:schematic_SEM}(b) supports a pair of degenerate TE$_{p=1}$ \emph{traveling wave}
WGM resonances with $Q = 3.4\times 10^5$ in the 852 nm wavelength band (blue curves of Fig.\ \ref{fig:WGM_properties}(a)). After placement of
the nanocrystal near the disk edge, the WGM resonance splits into an asymmetric pair of \emph{standing wave} resonances (red curves of Fig.\
\ref{fig:WGM_properties}(a)) formed from backscattering by the nanocrystal\cite{ref:Weiss, ref:Mazzei1}. The standing wave cavity modes
spatially lock to the position of the sub-wavelength nanocrystal, with the anti-node (node) of the lower (higher) frequency resonance aligned
with the nanocrystal.  As a result, the $Q$-factor of the long wavelength resonance is degraded to $Q=1.7\times 10^5$ due to scattering by the
nanocrystal, while the $Q$-factor of the shorter wavelength resonance remains relatively close to its unperturbed value.  The measured
mode-splitting and scattering loss induced by the nanocrystal is in good correspondence with a perturbative analysis (see Appendix), depending
upon similar cavity field properties as the dipole-cavity coherent coupling rate $g$ \cite{ref:Borselli2, ref:Mazzei1}.

\begin{figure}[tb]
\begin{center}
  \epsfig{figure=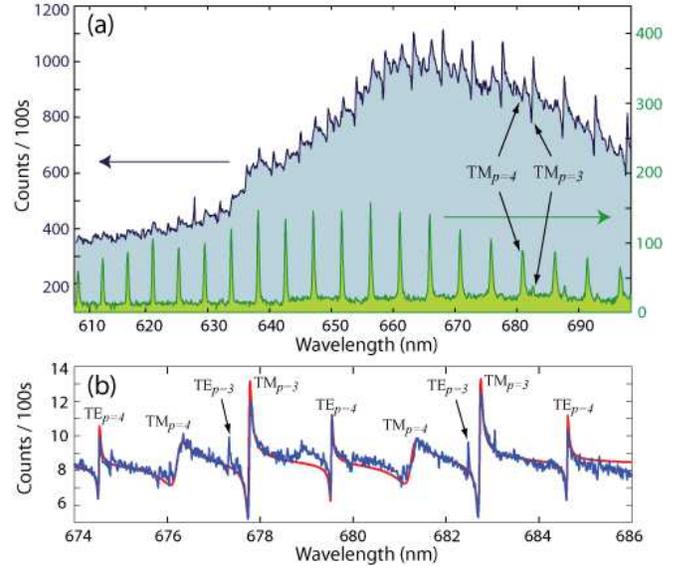, width=1\linewidth}
  \caption{(a) PL from the diamond nanocrystal placed on the microdisk in Fig.\ \ref{fig:schematic_SEM}(b), collected at
    room temperature through the near-field fiber taper (green) and far-field lens (blue). (b) High resolution ($\delta\lambda=20$ pm)
    PL spectrum of the lens-collected emission around $\lambda=680$ nm. The red curve is a fit using $S_{\text{lens}}$ from the text,
    generalized to include multiple decoupled cavity modes.}
\label{fig:NV_disk_data}
\end{center}
\end{figure}

Optical excitation of the assembled nanocrystal-microcavity system shows the striking influence of the microdisk cavity, and the resulting
interference between the different channels for nanocrystal emission. PL spectra collected in the far-field using the high-NA lens and the fiber
taper are shown in Fig.\ \ref{fig:NV_disk_data}. The fiber taper is evanescently coupled to the microdisk on the edge diametrically opposite
from the nanocrystal, and only collects light emitted into the cavity modes of the microdisk. In both collection geometries the \emph{envelope}
of the collected spectrum follows the broad NV$^-$ emission characteristic of Fig.\ \ref{fig:apparatus_taper_data}(b); we estimate that the
nanocrystal under study here contains less than $5$ NV centers.  The fiber taper PL spectrum consists primarily of regularly spaced Lorentzian
peaks corresponding to the microdisk WGM resonances.  The lens-collected PL, a high-resolution spectrum of which is shown in Fig.\
\ref{fig:NV_disk_data}(b), is instead punctuated by Fano-like resonances superimposed upon the broad background spectrum of the NV$^-$
transition.

For a nanocrystal positioned on the microdisk top surface, $150$ nm from the disk edge, FEM simulations show that in the $600$ nm wavelength
band the higher radial order ($p=3,4$) WGMs of the microdisk are most strongly coupled to the nanocrystal (Fig. \ref{fig:WGM_properties}(b,c)).
A perturbative analysis of the scattering loss induced by the nanocrystal (see Appendix) also indicates that the $Q$ values of the $p=1$-$4$
radial order modes, save the TM$_{p=4}$, should be limited by nanocrystal scattering.  From careful comparison of the measured emission spectra
with the calculated radiation-limited and nanocrystal-scattering-limited $Q$ values, the various families of cavity modes in the PL from the
nanocrystal can be identified, as indicated in Fig. \ref{fig:NV_disk_data}.  Although visible in the high-resolution lens-collected PL spectrum
(Fig. \ref{fig:NV_disk_data}(b)), the TE$_{p=3,4}$ modes are not faithfully reproduced due to the limited resolution of the spectrometer. Note
that doublet splitting is not expected for the these modes, owing to their lower-Q and large mode volume (see Appendix.) Absent from both the
lens and taper collected emission spectra are the highest $Q$, $p=1,2$ radial-order modes, a result of the low spectrometer resolution which
washes out narrow spectral features, and the tighter confinement of these modes inside the microdisk which weakens the coupling to the fiber
taper waveguide.

To understand the measured PL spectra from the two collection geometries further, we now turn to the model described by eq.\ (\ref{eq:spectra}).
For collection through the fiber taper, we observe no interference due to negligible overlap between the fiber taper mode and the radiation
modes of the nanocrystal ($\upepsilon_d/\upepsilon_c \sim 0$). The resulting emission spectrum is then dominated by the second term in eq.\
(\ref{eq:spectra}) resulting in Lorentzian peaks centered at each of the microdisk cavity resonance wavelengths as seen in the measured spectrum
of Fig.\ \ref{fig:NV_disk_data}.  For the far-field lens collection, $\upepsilon_d \approx \upepsilon_c$ when the cavity radiation is dominated
by scattering from the nanocrystal, as is estimated for all but the TM$_{p=4}$ modes. For a sub-wavelength nanocrystal in which the origin of
the dipole emission and the scattered cavity emission nearly coincide, one can show that $\phi_d - \phi_c = \pi/2$ (see Appendix), resulting in
a lens-collected power spectral density,
\begin{equation}\label{eq:spectrum_free}
S_{\text{lens}}(\omega) \propto \frac{1}{\gamma_p}\left|1 +
  i\sqrt{\frac{2g^2}{\kappa\gamma_s}}\frac{1}{1+i\Delta\omega/\kappa}\right|^2.
\end{equation}
\noindent Using the above relation for $S_{\text{lens}}$, a fit to the Fano resonances of the high resolution lens-collected spectrum is
performed (Fig. \ref{fig:NV_disk_data}(b)).  For the TM$_{p=3}$ modes, which are both spectrometer-resolved and scattering-limited, the fit
yields an estimate of $\kappa/2\pi = 15$ GHz and $F_o \approx 0.20$.  In the case of the spectrometer-resolved TM$_{p=4}$ modes we find
$\kappa/2\pi = 73$ GHz and $F_o \approx 0.020$; however, the reduced $\kappa F_{o}$ product for this mode family is consistent with a
radiation-limited $Q$-factor for which $\upepsilon_c < \upepsilon_d$.

Most directly, $F_{o}$ represents a ratio of the local density of states of the high-$Q$ cavity modes relative to that of the radiation modes,
as seen by an NV center embedded in the nanocrystal.  Fitting the data gives the ratio between $g^2$ and $\gamma_s$, from which one can obtain
$g$ by substituting the appropriate $\gamma_s$.  Of particular interest is the discrete ZPL transition of the NV$^{-}$ state whose linewidth can
approach its radiation-lifetime-limited value of $12$ MHz at cryogenic temperatures \cite{ref:Tamarat1}. Substituting $(\gamma_{s} = \gamma_{\text{zpl}})/2\pi
\approx 0.5$ MHz, $\kappa/2\pi = 15$ GHz and $F_{o}=0.2$ for the TM$_{p=3}$ resonances yields $g_{\text{zpl}}/2\pi \approx 28$ MHz.  This value
is roughly half that estimated from the simulated TM$_{p=3}$ mode overlap with the nanocrystal ($68$ MHz), and within a factor of 5 of the
estimated maximum value for this microdisk ($130$ MHz).  Factors contributing to the smaller observed $g_{\text{zpl}}$ include imperfect
mode-matching of the direct dipole emission and the nanocrystal-scattered cavity field, local field effects stemming from polarization of the
diamond host, and the NV position(s) within the relatively large nanocrystal, for which the presence of multiple NV centers with random dipole
orientations is likely to result in a lower average $g_{\text{zpl}}$.

These measurements provide an initial demonstration of deterministic placement of single diamond nanocrystals on optical microcavities. Using
this technique in future experiments to selectively manipulate smaller nanocrystals containing single, narrow linewidth NV centers, it should be
possible to fabricate a coupled NV-cavity system with Purcell enhanced emission exceeding unity.  Such a system would be an important step
towards the efficient optical readout and control of the electronic and nuclear states of an NV center, and the realization of NV center based
devices for quantum information processing \cite{ref:Duan1}, such as a quantum repeater.  More importantly, this method may be utilized to
integrate more complex diamond nanocrystal cavity QED systems, forming the basis of a chip-based quantum network.

\section*{Methods}

\begin{footnotesize}

  \noindent $\textbf{Microdisk fabrication}$ The microdisk optical cavities are SiO$_2$ disks, 20 $\mu$m in diameter,
  approximately $400$nm thick near the disk edge, and supported by a central SiO$2$ pillar $\sim 5$ $\mu$m in diameter. They were fabricated by thermally
  oxidizing Si microdisks in an oxygen purged furnace at a temperature of $1050^{\text{o}}$C. The template Si microdisks
  were fabricated from a silicon-on-insulator (SOI) wafer with a $2$ $\mu$m buried-oxide-layer and a p-doped $217$ nm Si device layer of resistivity $14$-$20$ $\Omega$cm.  The Si microdisk processing was as follows.  Electron beam lithography was
  used to define the microdisk pattern in a polymer electron beam resist (ZEP 520A).  To improve circularity and remove
  surface roughness, the patterned resist was reflowed at 160$^{\text{o}}$C \cite{ref:Borselli2}. A SF$_6$/C$_4$F$_8$ inductively
  coupled reactive ion etch transfers the resist pattern into the top Si layer.  An HF wet etch partially removes the
  underlying SiO$_2$ layer, resulting in Si microdisks supported by SiO$_2$ posts.

  \noindent $\textbf{Photoluminescence measurements}$ Diamond nanocrystals containing NVs are excited optically using a
  frequency-doubled YAG laser ($\lambda \approx 532$ nm).  A 50X (NA = 0.55) ultralong working distance microscope
  objective lens focuses the excitation beam to a 1.5 $\mu$m spot.  Typical excitation power is $250$ $\mu$W. A dichroic
  mirror on the backside of the objective separates NV photoluminescence from excitation power.  Photoluminescence is
  passed through a long wavelength pass filter (cut-off $\sim 540$ nm), and then directed with a flipper mirror to
  either a spectrometer (resolution $\sim 20$pm), or imaging optics and a CCD camera.

\end{footnotesize}

\section*{Acknowledgements}

\begin{footnotesize}

The authors thank Tom Johnson and Matt Borselli for setting up and calibrating the oxidation furnace used to oxidize the Si microdisks.  This
work was supported by DARPA and the Air Force Office of Scientific Research through AFOSR Contract No.\ FA9550-07-C-0030.

\end{footnotesize}

%\section*{Competing Financial Interests}

%\begin{footnotesize}

%The authors declare that they have no competing financial interests.

%\end{footnotesize}

\appendix

\section{Effective mode volume and coherent coupling rate}\label{sec:V_eff_and_g}

  In this work we use the commonly defined effective mode volume of a
  microcavity mode in terms of its peak electric field energy density:

  \begin{equation}
    V_{\text{eff}} = \frac{\int n^2 (\vrr) |\vE_c(\vrr)|^2 d^3r}{\text{max}\left[n^2 (\vrr) |\vE_c(\vrr)|^2\right]},
  \label{eq:Veff_defn}
  \end{equation}

  \noindent where $n(\vrr)$ is the refractive index profile of the cavity and $\vE_c(\vrr)$ is the electric field of the
  cavity mode.  For the whispering-gallery-modes of the microdisk, the quoted effective mode volumes, unless
  otherwise stated, are for standing-wave resonances, which are a factor of two smaller than their traveling-wave
  counterparts.  We use a factor $\eta(\vrr_{o})$ to relate the peak of the electric field energy density in the cavity
  to the electric field energy density at a particular position $\vrr_{o}$ ($\eta(\vrr_{o}) \ge 1$).  The per photon electric field amplitude at
  position $\vrr_{o}$ then relates to the effective mode volume as,

  \begin{equation}
  E_{c,\text{photon}}(\vrr_{o}) = \sqrt{\frac{\hbar \omega}{2\epsilon_o n^2(\vrr_o) \eta(\vrr_o) V_{\text{eff}}}},
  \label{eq:E_per_photon}
  \end{equation}

  \noindent from which the coherent coupling rate can be estimated as $g \equiv |\vmu \cdot \vE_{c,\text{photon}}(\vrr_{o})|/\hbar$.  The magnitude of the dipole moment of the electronic transition, $|\vmu|$, can be determined from the spontaneous emission rate in bulk
  material,

  \begin{equation}
  |\vmu|^2 = \frac{3 \pi^2 \hbar \epsilon_o   c^3 \gamma_{||}}{n_{\text{nc}} \omega^3}.
  \label{eq:mu}
  \end{equation}

\section{Mode-coupling from a sub-wavelength scatterer}

The modal coupling between clockwise ($cw$) and counterclockwise ($ccw$) traveling wave modes in a whispering-gallery-mode microcavity has been
observed experimentally and explained by many other authors, including those of Refs.
[\onlinecite{ref:Weiss,ref:Little3,ref:Kippenberg,ref:Gorodetsky1,ref:Borselli2}].  Here, we present a simple analysis of this coupling.
Maxwell's wave equation for the vector electric field in a microdisk structure is

\begin{equation}
\label{eq:Maxwell_repeat}
\vn{^2}\vE-\mu_{0}\Bigl(\epsilon^{0}+\delta\epsilon\Bigr)\frac{\partial^{2}\vE}{\partial
t^{2}}=0,
\end{equation}

\noindent where $\mu_{0}$ is the permeability of free space, $\epsilon^{0}$ is the dielectric function for the ideal (perfectly circular)
microdisk and $\delta\epsilon$ is the dielectric perturbation that is the source of mode coupling between the $cw$ and $ccw$ modes. Assuming a
harmonic time dependence, the modes of the ideal microdisk structure are
$\vE^{0}_{j}(\vecb{r},t)=\vE^{0}_{j}(\vecb{r})\text{exp}(i\omega_{j}t)$, and are solutions of eq. (\ref{eq:Maxwell_repeat}) with
$\delta\epsilon=0$.  Solutions to eq. (\ref{eq:Maxwell_repeat}) with $\delta\epsilon\neq0$ (i.e., modes of the perturbed structure) are written
as a sum of the unperturbed mode basis

\begin{equation}
\label{eq:Efield_solns}
\vE(\vecb{r},t)=e^{-i\omega_{0}t}\sum_{j}a_{j}(t)\vE^{0}_{j}(\vecb{r}).
\end{equation}

\noindent Plugging this equation into eq. (\ref{eq:Maxwell_repeat}), and utilizing mode orthogonality, we arrive at the coupled mode equations

\begin{align}
\label{eq:coupled_mode_1}
&\frac{da_{k}}{dt}+i\Delta\omega_{k}a_{k}(t)=i\sum_{j}\beta_{jk}a_{j}(t) \\
&\beta_{jk}=\frac{\omega_{0}}{2}\frac{\int\delta\epsilon\bigl(\vE^{0}_{k}(\vecb{r})\bigr)^{\ast} \cdot \vE^{0}_{j}(\vecb{r})d\vecb{r}}{\int\epsilon^{0}|\vE^{0}_{k}(\vecb{r})|^{2}d\vecb{r}},
\end{align}

\noindent where $\Delta\omega_k = \omega_k - \omega_0$. In deriving eq.\ (\ref{eq:coupled_mode_1}) we have  assumed that the mode amplitudes
change slowly relative to $\omega_0$, and that $|\omega_j - \omega_0| \ll \omega_0$. We have also ignored small "self term" corrections to the
eigenfrequencies. Reference \cite{ref:Borselli2} presents a functional form for $\beta$ in situations involving small surface roughness
perturbation. Under weak scattering conditions an assumption is made that only each pair of localized, degenerate $cw$ and $ccw$ WGMs of
azimuthal mode number $\pm m$ (with dominant polarization (TE or TM) and radial mode number $p$) are coupled by the disk perturbation
$\delta\epsilon$.  The coupled mode equations for these traveling wave modes then read as

\begin{equation}
\begin{split}
\label{eq:coupled_mode_2}
\frac{da_{cw}}{dt}&=-i\Delta{\omega}a_{cw}(t)+i|\beta|e^{i\xi}a_{ccw}(t) \\
\frac{da_{ccw}}{dt}&=-i\Delta{\omega}a_{ccw}(t)+i|\beta|e^{-i\xi}a_{cw}(t),
\end{split}
\end{equation}

\noindent with $\beta=|\beta|e^{i\xi}$ given by the integral in cylindrical coordinates $(\phi,\rho,z)$,

\begin{equation}
\label{eq:backscatter_2}
\beta=\frac{\omega_{0}}{2}\frac{\int\bigl(\int e^{+i2m\phi} \delta\epsilon (\phi,\rho,z)d\phi \bigr) \bigl(\vE^{0}_{cw}(\rho,z)\bigr)^2 \rho d\rho dz}{\int\epsilon^{0}|\vE^{0}_{cw}(\vecb{r})|^{2}d\vecb{r}}.
\end{equation}

\noindent where we have used the fact that $\vE^{0}_{cw}(\vrr) = \vE^{0}_{cw}(\rho,z)e^{+im\phi}$ and $\vE^{0}_{ccw}(\vrr) =
(\vE^{0}_{cw}(\vrr))^{*}$. For a sub-wavelength nanocrystal scatterer, the dielectric perturbation in eq. (\ref{eq:backscatter_2}) can be
approximated as

\begin{equation}
\delta\epsilon=\epsilon_{0}(n_{\text{nc}}^{2}-1)\delta^{(3)}(\vrr-\vrr_{\text{nc}})V_{\text{nc}},
\label{delta_ep}
\end{equation}

\noindent where $\epsilon_{0}$ is the free space permittivity, $n_{\text{nc}}$ is the refractive index of the diamond nanocrystal ($\sim 2.4$)
and $V_{\text{nc}}$ is the physical volume of the nanocrystal.  The mode-splitting in angular frequency is $2|\beta|$, and is proportional to
the center frequency $\omega_{0}$.  The normalized mode-splitting, in terms of the \emph{traveling-wave} mode effective volume (note this is
twice as large as the standing-wave mode volume), can then be written as

\begin{equation}
\label{eq:beta}
\frac{1}{Q_{\beta}} = \frac{2\left\vert\beta\right\vert}{\omega_{0}} = \frac{(n^2_{\text{nc}} - 1)V_{\text{nc}}}{\eta(\vrr_{\text{nc}}) V_{tw\text{, eff}}},
\end{equation}

\noindent where $\eta(\vrr_{\text{nc}})$ is the correction factor taking into account the reduced electric field strength at the position of the
nanocrystal.

We have performed measurements of the scattering effects on the TE$_{p=1}$ WGMs in the $850$ nm wavelength band of the $10$ $\mu$m radius,
SiO$_2$ microdisks studied in this work.  For the $200$ nm diameter nanocrystal placed on the microdisk top surface, $150$ nm from the disk
edge, the ratio of the mode-splitting to the center frequency of the TE$_{p=1}$ WGMs is measured to be $8.2 \times 10^{-6}$.  This corresponds
well with the theoretical value predicted by eq. (\ref{eq:beta}) of $2.2 \times 10^{-5}$ for a simulated \emph{traveling-wave} effective mode
volume $V_{tw\text{, eff}} = 86 (\lambda_o/n_{\text{SiO}_2})^3$ and $\eta=0.024$ evaluated at the center of the nanocrystal ($100$ nm above the
disk surface).

\section{Surface-scattering from a sub-wavelength scatterer}

Optical losses in whispering-gallery-mode resonators are often limited by refractive index perturbations, $\delta\epsilon$, on the cavity
surface. These index perturbations are sourced approximately by the unperturbed field solutions to create polarization currents,

\begin{equation}
\mathbf{J}=-i\omega\delta\epsilon\mathbf{E}.
\label{Pol_current}
\end{equation}

In analogy with microwave electronics, the polarization currents drive new electromagnetic fields which radiate into space. Optical losses due
to the perturbations can be calculated from the far field solutions setup by $\mathbf{J}$ \cite{Kuznetsov}.   The far field vector potential
sourced by $\mathbf{J}$ is given by\cite{Kuznetsov}

\begin{equation}
\mathbf{A}_{rad}(\mathbf{r})=\frac{\mu_{o}}{4\pi}\left( \frac{e^{-ik_{0}r}
}{r}\right) \int\vJ(\mathbf{r}^{\prime})e^{-ik_{0}\mathbf{\hat{r}}
\cdot\mathbf{r}^{\prime}}d\mathbf{r}^{\prime} \label{far_field},
\end{equation}

\noindent where $k_{0}$ is the wave vector in the surrounding air, and we have made the simplifying assumption that the microcavity does not
significantly modify the dipole radiation from its free space radiation pattern.

For a point-like nanocrystal scatterer with perturbation given by eq. (\ref{delta_ep}), and for source field corresponding to a standing-wave
WGM of the microdisk, the far field Poynting vector is given by

\begin{align}
\mathbf{S}_{rad} &=\mathbf{\hat{r}}\frac{\omega
k_{0}}{2\mu_{0}}\left\vert \mathbf{\hat{r}\times A}
_{rad}\right\vert ^{2}\\
 &= \mathbf{\hat{r}}\frac{\omega k_{0}^3 \left(n_{\text{nc}}^2-1\right)^{2} V_{\text{nc}}^{2} \epsilon_{0} \left\vert
\mathbf{E}_{c}(\vrr_{\text{nc}})\right\vert ^{2}}{32\pi^2}\frac{\left\vert \mathbf{\hat
{r}\times\hat{e}(\vrr_{\text{nc}}))}\right\vert ^{2}}{r^{2}}\notag,
\end{align}

\noindent where $\mathbf{E}_{c}(\vrr_{\text{nc}})$ is the cavity mode electric field and $\mathbf{\hat{e}}(\vrr_{\text{nc}})$ is the unit vector
representing the polarization of the cavity mode electric field at the position of the nanocrystal scatterer. The total (cycle-averaged) power
radiated, $P_{rad}$, far from the disk can be found by integrating the outward intensity over a large sphere,

\begin{align}
P_{rad}&=\int\left( \mathbf{S}\cdot\mathbf{\hat{r}}\right) r^{2}d\Omega\label{radiated_power}\\
&= \frac{\omega k_{0}^3 \left(n_{\text{nc}}^2-1\right)^{2} V_{\text{nc}}^{2} \epsilon_{0} \left\vert
\mathbf{E}_{c}(\vrr_{\text{nc}})\right\vert ^{2}}{32\pi^2} \int \left\vert \mathbf{\hat
{r}\times\hat{e}(\vrr_{\text{nc}}))}\right\vert ^{2}d\Omega.\notag
\end{align}

\noindent For the sphere centered about the nanocrystal scatterer, $\int \left\vert \mathbf{\hat {r}\times\hat{e}(\vrr_{\text{nc}}))}\right\vert
^{2}d\Omega = 8\pi/3$.  The quality factor of a cavity is given by $Q=\omega U_{c}/P_{rad}$, where $U_{c}=$ $\frac{1}
{2}\int\epsilon^{0}(\mathbf{r)}\left\vert \mathbf{E}_{c}\right\vert ^{2} d\mathbf{r}$ is the cycle-averaged stored energy in the cavity.  The
cavity energy can also be related to the effective mode volume through eq. (\ref{eq:Veff_defn}).  Combining all of these relations, we can
rewrite eq. (\ref{radiated_power}) as a scattering quality factor

\begin{equation}
\label{eq:Qs}
Q_{s} = \frac{3 \lambda_{o}^{3} \eta(\vrr_{\text{nc}}) V_{\text{eff}} }{4 \pi^2 (n_{\text{nc}}^2 - 1)^2 V_{\text{nc}}^2}.
\end{equation}

\noindent Estimates of the scattering quality factor for the various WGMs studied in this work are presented in Tables \ref{tab:params_637} and
\ref{tab:params_850}, where the nanocrystal has a measured diameter of $200$ nm and the correction factor $\eta(\vrr_{\text{nc}})$ is evaluated
at the center of the nanocrystal (approximately $100$ nm above the disk surface).

\section{Finite-element-method simulations of SiO$_2$ microdisk modes}

In Table \ref{tab:params_637} and \ref{tab:params_850} we present the results of finite-element-method (FEM) simulations of the $10$ $\mu$m
radius, SiO$_2$ microdisks used in this study at resonant wavelengths in the $650$ nm and $850$ nm wavelength bands, respectively.  The
effective index of each cavity mode is calculated from the approximate relation, $n_{\text{eff}} \approx m \lambda_o/2\pi R_o$, where $m$ is the
azimuthal mode number of the WGM and $R_o = 10$ $\mu$m is the physical radius of the microdisk.  The correction factors, $\eta_{s}$ and
$\eta_{nc}$, correspond to the electric field energy density evaluated at the radial position of the nanocrystal ($\sim 150$ nm from the disk
edge) and vertical position at the surface of the microdisk and at the center of the nanocrystal ($100$ nm above the disk surface),
respectively.  The surface-scattering-limited $Q$-factor is estimated from eq. (\ref{eq:Qs}) using $\eta_{nc}$.

\renewcommand{\arraystretch}{1.0}
\renewcommand{\extrarowheight}{2pt}
\begin{table}
\caption{\textbf{Calculated mode parameters in the $\lambda \sim 600$ nm band.}} \label{tab:params_637}
\begin{center}
\begin{tabular}{llllllll}
%\begin{tabularx}{\linewidth}{YYYYYYYYYY}
\hline
\hline
Mode \hspace{8pt} & $Q_{\text{rad}}$ \hspace{22pt} & $V_{\text{eff}}$ \hspace{24pt}     & $\eta_{s}$ \hspace{2pt}  & $\eta_{\text{nc}}$ & $Q_{ss}$ & $m$ & $n_{\text{eff}}$ \\
                  &                             & $(\lambda_{o}/n_{\text{SiO}_{2}})^3$ &                         &                  &         &               &   \\
%\hhline{|=:=:=:=:=:=:=:=:=|}
\hline
TE$_{p=1}$ & $10^{13}$ & 69                           & 0.057     & 0.013  & $3.6 \times 10^4$ & 125 & 1.27\\
TM$_{p=1}$ & $10^{11}$ & 82                           & 0.061     & 0.021  & $2.7 \times 10^4$  & 122 & 1.24\\
TE$_{p=2}$ & $10^{10}$ & 97                           & 0.11      & 0.026  & $2.5 \times 10^4$ & 121 & 1.23\\
TM$_{p=2}$ & $10^{8}$ & 106                           & 0.21      & 0.069  & $1.0 \times 10^4$ & 118 & 1.20\\
TE$_{p=3}$ & $1.7 \times 10^{7}$ & 103                 & 0.12     & 0.028  & $2.5 \times 10^4$ & 115 & 1.17\\
TM$_{p=3}$ & $8.6 \times 10^{5}$ & 106                 & 0.24     & 0.079  & $9.0 \times 10^3$ & 113 & 1.15\\
TE$_{p=4}$ & $1.7 \times 10^{5}$ & 109                 & 0.11     & 0.026  & $2.9 \times 10^4$ & 111 & 1.13\\
TM$_{p=4}$ & $1.7 \times 10^{4}$ & 111                & 0.23    & 0.077  & $1.0 \times 10^4$ & 109 & 1.11\\
\hline
\hline
%\end{tabularx}
\end{tabular}
\end{center}
\end{table}
\renewcommand{\arraystretch}{1.0}
\renewcommand{\extrarowheight}{0pt}

\renewcommand{\arraystretch}{1.0}
\renewcommand{\extrarowheight}{2pt}
\begin{table}
\caption{\textbf{Calculated mode parameters in the $\lambda \sim 850$ nm band.}} \label{tab:params_850}
\begin{center}
\begin{tabular}{llllllll}
%\begin{tabularx}{\linewidth}{YYYYYYYYYY}
\hline
\hline
Mode \hspace{8pt} & $Q_{\text{rad}}$ \hspace{22pt} & $V_{\text{eff}}$ \hspace{24pt}     & $\eta_{s}$ \hspace{2pt}  & $\eta_{\text{nc}}$ & $Q_{ss}$ & $n_{\text{eff}}$ & $m$ \\
                  &                             & $(\lambda_{o}/n_{\text{SiO}_{2}})^3$ &                         &                  &         &               &     \\
%\hhline{|=:=:=:=:=:=:=:=:=|}
\hline
TE$_{p=1}$ & $4.5 \times 10^{8}$ & 43                & 0.073     & 0.024  & $1.4 \times 10^5$ & 1.27 & 93 \\
TM$_{p=1}$ & $2.2 \times 10^{6}$ & 51                & 0.086     & 0.040  & $8.4 \times 10^4$  & 1.22 & 87 \\
\hline
\hline
%\end{tabularx}
\end{tabular}
\end{center}
\end{table}
\renewcommand{\arraystretch}{1.0}
\renewcommand{\extrarowheight}{0pt}

  \section{Nanocrystal-cavity spectrum calculation}

  The measured spectrum for the nanocrystal-microdisk system is given by

  \begin{equation}
    S_{\text{lens}}(\omega) = \text{Re}\left[\int_{0}^{\infty}dt\int_{0}^{\infty}dt' e^{i\omega(t-t')}
      \left\langle\hat{\mathbf{E}}_o^\dagger(\mathbf{r},t)\cdot\hat{\mathbf{E}}_o(\mathbf{r},t')\right\rangle\right]
  \label{eq:PSD_lens_derivation}
  \end{equation}

  \noindent where $\hat{\mathbf{E}}_o(\mathbf{r},t)$ is the component of the radiated nanocrystal-microdisk field,
  $\hat{\mathbf{E}}(\mathbf{r},t)$, collected by the microscope objective or fiber taper collection optic. $\hat{\mathbf{E}}(\mathbf{r},t)$ is given as
  a function of the system variables $\hat{a}(t)$ and $\hat{\sigma}_-(t)$ by eq.\ (1) in the manuscript text.  To
  predict the spectrum of a spontaneous emission event, we assume that at $t = 0$, the dipole is prepared in its excited state and the microdisk contains no photons, and follow the techniques given in
  Refs.\ \cite{ref:Carmichael5,ref:Carmichael4} to calculate $S_{\text{lens}}(\omega)$ from $\langle\hat{a}(t)\rangle$ and
  $\langle\hat{\sigma}_-(t)\rangle$, as a function of the coupling and loss parameters of the system,
  $g,\kappa, \gamma_s, \gamma_p$, and detuning $\Delta\omega$ between the dipole and microdisk resonances.  An outline of this method is given
  below.

  The system Hamiltonian, which describes the dipole-cavity interaction in absence of decoherence, is given by
  \begin{equation}
  \hat H_S = \hbar\omega_c \hat{a}^\dagger \hat a + \hbar\omega_d\hat\sigma_+\hat\sigma_- + i \hbar g(\hat a^\dagger\hat\sigma_- - \hat a \hat \sigma_+),
  \end{equation}
where $\omega_d$ and $\omega_c$ are the dipole transition and cavity mode frequencies, $\hat\sigma_- = \hat \sigma_+^\dagger$ and $\hat a$ are
the dipole and cavity excitation annihilation operators, respectively, and  $g$ is the coherent dipole-cavity coupling rate defined in
\ref{sec:V_eff_and_g}. Coupling between the system variables and the measured radiation field is taken into account using the quantum master
equation for the system density matrix $\hat\rho$:
\begin{equation}
\frac{d{\hat \rho}}{dt} = \frac{1}{i\hbar}[\hat H_S,\hat \rho] + (\hat L_d + \hat L_c +\hat  L_p)\hat \rho
\end{equation}
where $\hat L_{a,c,p}$ are Lindblad operators given by:
\begin{align}
\hat L_d\hat \rho &= \frac{\gamma_s}{2}(2\hat \sigma_-\hat \rho\hat \sigma_+ - \hat \sigma_+\hat \sigma_-\hat \rho - \hat \rho\hat \sigma_+\hat \sigma_-) \\
\hat L_c\hat \rho &= \kappa(2 \hat a \hat \rho\hat  a^\dagger -\hat  a^\dagger\hat  a \hat \rho - \hat \rho \hat a^\dagger \hat a) \\
\hat L_p \hat \rho &= \frac{\gamma_p}{2}(\hat \sigma_z\hat \rho\hat \sigma_z - \hat \rho)
\end{align}
with $\hat\sigma_z = \hat\sigma_+\hat\sigma_- - \hat\sigma_-\hat\sigma_+$.  $\hat L_d$ ($\hat L_c$) describes decay of the dipole (cavity)
excitation into radiation modes, at energy decay rate $\gamma_s$ ($2\kappa$). $\hat L_p$ represents decoherence of the dipole excitation, in the
form of pure non-radiative dephasing at rate $\gamma_p$.

Equations of motion for the classical expectation values of $\hat a$ and $\hat \sigma_-$ can found from the quantum master equation, using the
relation $\frac{d}{dt}{\langle \hat O \rangle} = \text{tr}\left( \frac{d\hat\rho}{dt} \hat O \right)$:
\begin{align}
\frac{d}{dt}&\langle \hat a \rangle = -\left(i\omega_c + \kappa\right)\langle \hat a \rangle + g\langle\hat  \sigma_- \rangle   \label{eq:eom_a} \\
\frac{d}{dt}&\langle \hat \sigma_- \rangle = -\left(i\omega_d + \frac{\gamma_s}{2} + \gamma_p \right)\langle\hat\sigma_- \rangle + g\langle \hat a ~ \hat\sigma_z \rangle. \label{eq:eom_sigma}
\end{align}
Note that in absence of an external driving field, when a dipole is initially prepared in its excited state in a cavity with zero photons,
$\langle \hat a ~ \hat\sigma_z \rangle = - \langle \hat a \rangle$ for all time $t$.  These equations of motion can be used calculate the
two-time correlation functions required to determine the optical spectrum defined by eq.\ (\ref{eq:PSD_lens_derivation}). This is accomplished
using the quantum regression theorem, which, from eqs.\ (\ref{eq:eom_a}) and (\ref{eq:eom_sigma}), allows us to write, for $t' > t$,
\begin{align}
\frac{d}{dt'}\langle \hat a^\dagger(t)\hat a(t')\rangle = -(i\omega_c + \kappa)&\langle\hat a^\dagger(t) \hat a(t')\rangle \\
+&g\langle\hat  a^\dagger(t)\hat \sigma_-(t')\rangle \notag
\end{align}
\begin{align}
\frac{d}{dt'}\langle \hat a^\dagger(t)\hat \sigma_-(t')\rangle = -(i\omega_d + \frac{\gamma_s}{2} + &\gamma_p)\langle\hat  a^\dagger(t)
\hat \sigma_-(t')\rangle \\
 -& g\langle \hat a^\dagger(t)\hat a(t')\rangle\notag
\end{align}
 \begin{align}
\frac{d}{dt'}\langle\hat \sigma_+(t)\hat a(t')\rangle &= -(i\omega_c + \kappa)\langle\hat \sigma_+(t)\hat a(t')\rangle +
g\langle\hat \sigma_+(t)\hat \sigma_-(t')\rangle
\end{align}
\begin{align}
\frac{d}{dt'}\langle \hat \sigma_+(t)\hat \sigma_-(t')\rangle = -(i\omega_d + \frac{\gamma_s}{2} + &\gamma_p)\langle\hat  \sigma_+(t)
\hat \sigma_-(t')\rangle \\
- &g\langle \hat \sigma_+(t)\hat a(t')\rangle\notag.
\end{align}
Writing these equations, as well as analogous equations of motion for the special case $t = t'$, in the frequency domain, a closed set of
algebraic equations for the various contributions to $S_{\text{lens}}(\omega)$ is obtained.  Note that we have not assumed that the spontaneous
emission process is stationary. In the room-temperature limit,  $\gamma_p \gg \left[|\omega-\omega_{d,c}|, g, \kappa, \gamma_s\right]$, we find
\begin{align}
C_{cd}(\omega) &=  \mathbf F\left[\langle \hat a^\dagger(t)\hat\sigma_-(t')\rangle\right] = 0 \label{eq:C_cd}\\
C_{cc}(\omega) &=  \mathbf F\left[\langle \hat a^\dagger(t)\hat a(t')\rangle\right] = \frac{g^2}{\gamma_p\gamma_s\kappa}\frac{1}{i\Delta\omega + \kappa}\\
C_{dd}(\omega) &=  \mathbf F\left[\langle \hat \sigma_+(t)\hat\sigma_-(t')\rangle\right] =\frac{1}{\gamma_p\gamma_s}\\
C_{dc}(\omega) &=  \mathbf F\left[\langle \hat \sigma_+(t)\hat a(t')\rangle\right] =\frac{g}{\gamma_p\gamma_s}\frac{1}{i\Delta\omega+\kappa}\label{eq:C_dc}
\end{align}
where $\mathbf F\left[g(t,t')\right] = \int_{0}^{\infty}dt\int_{0}^{\infty}dt' e^{i\omega(t-t')}g(t,t')$ and $\Delta\omega = \omega - \omega_c$,
and we set $\omega_d = \omega_c$. From eqs.\ (\ref{eq:C_cd} - \ref{eq:C_dc}), eq.\ (\ref{eq:PSD_lens_derivation}), and eq.\ (1) in the text, we
arrive at the expression for $S(\omega)$ given by eq.\ (2) in the text.

\section{Dipole-cavity radiation phase lag}
Here we derive the phase difference between the radiation directly emitted from a dipole source (e.g., a diamond NV center) and the scattered
radiation from a cavity resonance driven by the same dipole source. In particular we consider the situation explored in the main text in which
the cavity scattering site is subwavelength and coincident with the dipole site (e.g., a diamond nanocrystal). We will show that the phase
difference, $\delta\phi = \phi_d - \phi_c$, is simply due to the $\pi/2$ phase shift between driving source and  local cavity field when driven
on resonance.  We begin with Maxwell's equations for a microcavity defined by $\epsilon_c(\vrr)$, a dielectric nanocrystal perturbation
$\Delta\epsilon_n(\vrr)$ which couples the microcavity mode to radiation modes, and a dipole source term $\mathbf{J}_s(\vrr,t)$:
\begin{equation}\label{eq:Maxwells_eqn}
\vn\times\vn\times\vE(\vrr,t)+\frac{\partial^2}{\partial t^2}\epsilon_c(\vrr)\vE(\vrr,t) = -\frac{\partial^2}{\partial t^2}\Delta\epsilon_n(\vrr)\vE(\vrr,t)
- \frac{\partial}{\partial t}\vJ_s(\vrr,t).
\end{equation}
We have assumed that the microcavity is non magnetic, and that the vacuum dielectric and magnetic permittivities are unity.

 To derive temporal equations of motion, we expand $\vE$ in terms of the cavity and radiation modes,
\begin{equation}\label{eq:mode_expansion}
\vE(\vrr,t) = c(t)\ve_c(\vrr)e^{-i\omega_c t} + \sum_j r_j(t)\ve_j(\vrr)e^{-i\omega_j t},
\end{equation}
where $\ve_{c,j}(\vrr)$ are solutions to Maxwell's equations in absence of the nanocrystal and external source,
\begin{equation}
\vn\times\vn\times\ve_{c,j}(\vrr)-\omega^2_{c,j}~\epsilon_c(\vrr)\ve_{c,j}(\vrr) = 0, \label{eq:Maxwell_harmonic}
\end{equation}
and $\omega_{j,c}$ are real. It follows that $\ve_{c,j}(\vrr)$ are orthogonal, and we choose to normalize the fields as follows:
\begin{align}
\langle c | \e_c(\vrr) | j \rangle &= \int d\vrr~ \e_c(\vrr)~\ve_c^*(\vrr)\cdot\ve_j(\vrr) = 0 \label{eq:orthog_1},\\
\langle i | \e_c(\vrr) | j \rangle &= \int d\vrr~ \e_c(\vrr)~\ve_i^*(\vrr)\cdot\ve_j(\vrr) = \delta_{ij}\label{eq:orthog_2}, \\
\langle c | \e_c(\vrr) | c \rangle &= \int d\vrr~ \e_c(\vrr)~\ve_c^*(\vrr)\cdot\ve_j(\vrr) = 1 \label{eq:orthog_3}.
\end{align}
 Note that in this mode expansion, $\ve_c(\vrr)$ represents a truly bound mode
of the cavity: in absence of the nanocrystal or other perturbation, excitations of this mode have an infinite lifetime.

Substituting eq.\ (\ref{eq:mode_expansion}) into eq.\ (\ref{eq:Maxwells_eqn}), and applying eqs.\ (\ref{eq:Maxwell_harmonic} -
\ref{eq:orthog_3}), for a harmonic driving term $\vJ_s(\vrr,t) = \vJ_o(\vrr)e^{-i\omega_s t}$, the following coupled mode equations can be
obtained:
\begin{align}
\frac{d}{dt}c(t) &= \sum_j i\frac{\omega_j^2}{2\omega_c}\upkappa_{cj}r_j(t)e^{-i(\omega_j - \omega_c)t}+\frac{\omega_s}{2\omega_c}s_c e^{-i(\omega_s-\omega_c)t} \label{eq:cm_1},\\
\frac{d}{dt}r_j(t) &= i\frac{\omega_c^2}{2\omega_j}\upkappa_{jc}c(t)e^{-i(\omega_c - \omega_j)t}+\frac{\omega_s}{2\omega_j}s_je^{-i(\omega_s-\omega_j)t}\label{eq:cm_2},
\end{align}
where,
\begin{equation}
\upkappa_{cj} = \upkappa_{jc}^* = \langle c | \Delta\e_n| j \rangle, \\
\end{equation}
\begin{equation}
s_{c,j} = -\langle c,j | \vJ_o \rangle.
\end{equation}
In deriving eqs. (\ref{eq:cm_1})-(\ref{eq:cm_2}), we have ignored ``self-terms" which result in corrections to the eigenfrequencies, and have
assumed that the mode amplitudes vary slowly compared to the optical frequencies. We have also assumed that the field $\ve_c(\vrr)$ is
predominantly confined within the cavity, and that fields $\ve_j(\vrr)$ are predominantly excluded from the cavity.  Upon integrating eq.\
(\ref{eq:cm_2}) directly, and substituting the result into eq.\ (\ref{eq:cm_1}), we arrive at a differential equation for $c(t)$.  Assuming that
the radiation modes form a continuum with slowly varying density of states, $\rho_o$, and coupling coefficients, $\upkappa_c \equiv
\upkappa_{cj}$, and that coupling from the radiation modes \emph{into} the cavity is negligible, a simplified equation of motion for $c(t)$ is
obtained:
\begin{equation}
\frac{d}{dt}\tilde{c}(t) = \left(-\kappa + i(\omega_s - \omega_c)\right)\tilde{c}(t) + \frac{\omega_s}{2\omega_c}s_c\label{eq:eom_c},
\end{equation}
where $\tilde{c}(t) = e^{i(\omega_s - \omega_c)t}c(t)$, and $2\kappa = \pi\rho_o\omega_c^2|\upkappa_{c}|^2/2$ is the total energy decay rate of
the cavity mode into the radiation mode continuum, due to scattering from the nanocrystal. Averaging over long times, only radiation into the
mode whose frequency is resonant with the source frequency will be appreciable; the equation of motion for the mode amplitude, $r_s \equiv
r_{j=s}$, is
\begin{equation}
\frac{d}{dt}\tilde{r}_s(t) = \frac{s_s}{2} + \frac{s_c}{4}\frac{i~\omega_c\upkappa_{c}^*}{i(\omega_c-\omega_s)+\kappa} \label{eq:eom_rs},
\end{equation}
where $\tilde{r}_j(t) = r_j(t) e^{i(\omega_s-\omega_j)t}$, and we have used the steady state solution for $\tilde{c}(t)$ from eq.\
(\ref{eq:eom_c}).

The Fano nature of radiation from the nanocrystal-cavity system can seen in the righthand side eq.\ (\ref{eq:eom_rs}): the first term represents
emission from the source ``directly" into the radiation modes, while the second term represents ``indirect" emission from the source into the
radiation modes via the cavity mode.  Note that eq.\ (\ref{eq:eom_rs}) has the same form as eq.\ (2) in the text, the primary difference lying
in the definitions of the various coupling coefficients.  Here, the relative phase of the coupling coefficients is well defined by Maxwell's
equation, and can easily be determined in the case of a subwavelength scatter such as a nanocrystal.

Assuming that $\vJ_o(\vrr) = \mathbf{j}_o\delta^{(3)}(\vrr-\vrr_o)$ and $\Delta\e_n(\vrr) = \delta\e_n\delta^{(3)}(\vrr-\vrr_o)$, the coupling
coefficients are given by,
\begin{align}
\upkappa_c &= \ve_c^*(\vrr_o)\cdot\ve_s(\vrr_o)\delta\e_n = e^{-i(\phi_c-\phi_s)}\left|\ve_c^*(\vrr_o)\cdot\ve_s(\vrr_o)\delta\e_n\right|, \\
s_c &= -\ve_c^*(\vrr_o)\cdot\mathbf{j}_o = e^{-i(\phi_c-\phi_J+\pi)}\left|\ve_c^*(\vrr_o)\cdot\mathbf{j}_o\right|,\\
s_s &= -\ve_s^*(\vrr_o)\cdot\mathbf{j}_o = e^{-i(\phi_s-\phi_J+\pi)}\left|\ve_s^*(\vrr_o)\cdot\mathbf{j}_o\right|,
\end{align}
and eq.\ (\ref{eq:eom_rs}), with the relative phases included explicitly, is
\begin{equation}
\frac{d}{dt}\tilde{r}_s(t) = e^{i(\phi_J-\phi_s+\pi)}\left[\frac{|s_s|}{2}+\frac{|s_c|}{4}\frac{i~\omega_c|\upkappa_{c}^*|}{i(\omega_c-\omega_s)+\kappa}\right]\label{eq:eom_rs_phase}.
\end{equation}
The above equation indicates that for the system of interest, when the cavity and source are on resonance, ``direct" and ``indirect"
contributions to the radiation field are $\pi/2$ out of phase.

\end{document}